\documentclass[pra,twocolumn]{revtex4}

\usepackage{graphicx}
\usepackage{hyperref}
\usepackage{amssymb,amsmath}
\usepackage{epstopdf}

\begin{document}

\title{Effect of the heralding detector properties \\on the conditional generation of single-photon states}

\author{V. D'Auria$^{2}$, O. Morin$^{1}$, C. Fabre$^{1}$, and J. Laurat$^{1}$}

\affiliation{$^{1}$Laboratoire Kastler Brossel, Universit\'{e}
Pierre et Marie Curie, Ecole Normale Sup\'{e}rieure, CNRS, 4 place
Jussieu, 75252 Paris Cedex 05, France\\
$^{2}$Laboratoire de Physique de la Mati\`{e}re Condens\'{e}e, CNRS UMR 7336,
Universit\'{e} de Nice - Sophia Antipolis,
 Parc Valrose, 06108 Nice Cedex 2, France}

\begin{abstract}
Single-photons play an important role in emerging quantum technologies and information processing. An efficient generation technique consists in preparing such states via a conditional measurement on photon-number correlated beams: the detection of a single-photon on one of the beam can herald the generation of a single-photon state on the other one. Such scheme strongly depends on the heralding detector properties, such as its quantum efficiency, noise or photon-number resolution ability. These parameters affect the preparation rate and the fidelity of the generated state. After reviewing the theoretical description of optical detectors and conditional measurements, and how both are here connected, we evaluate the effects of these properties and compare two kind of devices, a conventional on/off detector and a two-channel detector with photon-number resolution ability. 
\end{abstract}

\maketitle

\section{Introduction}

A driving component for the development of diverse quantum information applications is the ability to efficiently engineer non-classical states of light \cite{Illuminati}. In particular, a great number of communication or computing protocols require single-photon states \cite{OBrien}. The control of single quantum objects provides a way to generate such states \cite{Lounis}. Another procedure, probabilistic but heralded, consists in measuring one mode of an entangled state, which results in projecting the other mode according to the measurement result: in the ideal case of twin beams, the detection of a single-photon in one beam projects the other one into a single-photon state \cite{Mandel}. The physical realization of this scheme critically relies on the performances of the twin-photon generator but also of the single-photon detector used to herald the preparation.  

Indeed, realistic detectors do not perfectly resolve photon numbers, have limited quantum efficiency and can be noisy \cite{Hadfield}. These properties directly translate into the fidelity of the generated state and its preparation rate. The goal of this paper is to evaluate how these properties impact the conditional preparation of single-photon states. We will consider the typical case of twin beams produced by two-mode spontaneous parametric down conversion. To illustrate the effect of the heralding detector properties, we will analyze two different detectors, a conventional APD and a single-loop time-multiplexed detector. The present work provides a detailed analysis of the experimental characterization of these detectors \cite{Dauria} and of the modeling of the single-photon state preparation using such detectors.

The paper is organized as follows. In section \ref{section2}, we first recall the general description of optical detectors by positive operator valued measure (POVM) and describe the conditional measurement scheme based on entangled beams. Given the detector POVM, we derive the general expression of the generated state. We illustrate this section by considering two specific detectors. Section \ref{section3} is then devoted to the experimental characterization of such detectors via quantum detector tomography. In Section \ref{section4}, by numerically simulating the entanglement resource, we predict the properties of the generated state that would be obtained if such detectors are used. We compare the two devices. We finally discuss the dependence of the generated state in the general case obtained by using the theoretical expressions of the POVM elements. This study enables us to determine the fidelity and preparation rate for the two heralding detectors, as a function of the noise background, quantum efficiency and entanglement resource.  

\section{Theoretical description of single-photon detectors and quantum state engineering}
\label{section2}

Quantum state engineering is an interplay between the entangled resource and the measurement performed by the heralding detector. In this section, we remind how the detector properties can be described in the most general case by a positive-operator valued measure and we provide the general expression of the state prepared by a measurement performed on one beam of the two correlated ones generated by spontaneous parametric dow-conversion. Two particular detectors are considered. 

\subsection{Description of a detector by a POVM}
A detector can be fully described by its positive operator valued measure \cite{Helstrom}, i.e. by a set of semipositive hermitian \{$\hat{\Pi}_{n}$\} associated to each of its $N$ possible outcomes labeled by integer ``\emph{n}''. These operators sum up to the identity, $\sum_{n=0}^{N-1}\hat{\Pi}_{n}=\hat{1}$. They enable to make predictions on the measurement outcomes. For any impinging states $\hat{\rho}$  on the detector, the probability $p_{n}(\hat{\rho})$ of obtaining the outcome ``\emph{n}'' is given by: 
\begin{equation}
p_{n}(\hat{\rho})=\textrm{Tr}[\hat{\rho}\,\hat{\Pi}_{n}] \label{Def}
\end{equation}
where $\textrm{Tr}$ stands for the trace.

In the following, we will consider the case of photon counters, which are phase-insensitive. In this specific case, the outcomes ``\emph{n}'' correspond to ``\emph{n clicks}''. For ideal detectors with perfect photon number resolution ability, the number of clicks directly gives the number of impinging photons. Therefore, the corresponding POVM is the set of projectors on the photon-number states:  $\hat{\Pi}_n=|n\rangle\langle n| $. 

However, realistic detectors are affected by non-unit quantum efficiency, imperfect counting ability and intrinsic noise. The general expression for $\hat{\Pi}_{n}$ can then be written as a sum of projectors:
\begin{equation}
\hat{\Pi}_{n}=\sum_{k=0}^{\infty} r_{k,n} |k\rangle\langle k|.  \label{POVM}
\end{equation}
Let us note that the coefficient $r_{k,n}$ can be read as:
\begin{equation}
r_{k,n}=\textrm{Tr}[\hat{\Pi}_{n}\,|k\rangle\langle k|]=P(n|k). 
\label{condprob}
\end{equation}
It thus corresponds to the conditional probability $P(n|k)$ to obtain the outcome ``\emph{n}'' giving an impinging Fock state $|k\rangle$. This expression is very convenient to determine the POVM coefficients of a given detector implementation.

\subsection{General expression of the POVM for two experimental cases}

We will consider here the case of two single-photon counters: a conventional $\emph{on/off}$ avalanche photodiode (APD) and a home-made single-loop time-multiplexed detector (TMD). We give the expression of the POVM elements in the general case, including non-unit quantum efficiency $\eta$ and dark noise with Poissonian statistics, which corresponds to many practical situations.

\subsubsection{Including non-unit quantum efficiency}

A conventional APD has a limited counting ability as it provides a binary outcome: it is only able to tell whether photons (whatever their number) have been detected ($\textrm{\emph{on}}$, one click) or not ($\textrm{\emph{off}}$, no click). In the ideal case, the two corresponding POVM elements are given by $\hat{\Pi}_{\textrm{\emph{off}}}=|0\rangle\langle 0|$ and $\hat{\Pi}_{\textrm{\emph{on}}}=\hat{1}-|0\rangle\langle 0|$. In the case of non-unit quantum efficiency $\eta$, the POVM for the APD detector can be written as \cite{Ferraro}:
\begin{eqnarray}
\left\{
    \begin{array}{l}
        \hat{\Pi}_{\textrm{\emph{off}}}(\eta)=\sum_{k=0}^{\infty} r_{k,\textrm{\emph{off}}}(\eta) |k\rangle\langle k|=\sum_{k=0}^{\infty}(1-\eta)^k |k\rangle\langle k|  \\
        \\
        \hat{\Pi}_{\textrm{\emph{on}}}(\eta)=\sum_{k=0}^{\infty} r_{k,\textrm{\emph{on}}}(\eta) |k\rangle\langle k|=\hat{1}-\hat{\Pi}_{\textrm{\emph{off}}}(\eta,\nu).
    \end{array}\right.
    \label{POVMAPD}
    \end{eqnarray}

As stated by equation (\ref{condprob}), the coefficient $r_{k,\textrm{\emph{off}}}$ is given by the expectation value of the operator $\hat{\Pi}_{\textrm{\emph{off}}}$ for an impinging Fock state $|k\rangle$. In the absence of noise, this value is equal to the probability of not detecting $k$ photons for a quantum efficiency $\eta$: it is indeed equal to the probability to not detect a photon, $1-\eta$, to the power of $k$. 

\begin{figure*}
\begin{center}
\includegraphics[width=1.6\columnwidth]{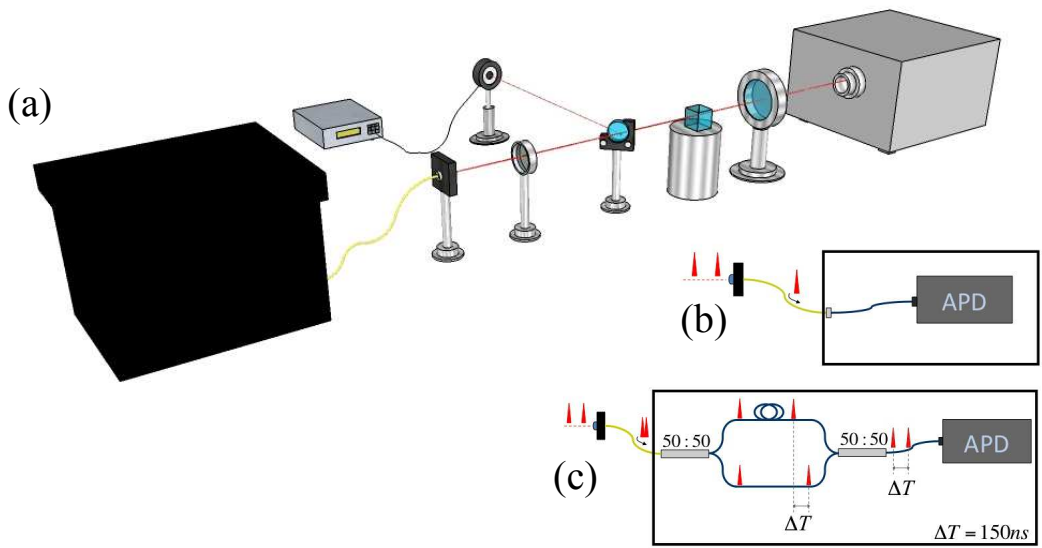}
\caption{Experimental quantum detector tomography. (a) The detector, shown here as a black-box, is probed with coherent pulses coupled into a single-mode fiber. A fraction of the light is tapped and measured with a calibrated power-meter to adjust the amplitude of the probe. Two detectors are characterized: (b) a single-photon counting module (APD) and (c) a two-channel time-multiplexed detector.}
\label{figure1}
\end{center}
\end{figure*}

The second detector we consider is a single-loop time-multiplexed detector with photon-number resolution ability (Fig. \ref{figure1}c). Indeed, it is able to provide an additional information by distributing the light to be detected in two temporal bins. The incoming photons are sent to a beamsplitter (reflectivity $\emph{R}$ in intensity) whose outputs are delayed by a time interval $\Delta T$ and recombined before being detected by an APD \cite{Achilles}. If two photons impinge on the detector, they are split toward different paths with probability $2R(1-R)$ and, in this case, they are detected by the APD as two subsequent events. Three outcomes are thus possible: $\emph{no click}$, $\emph{1 click}$ and $\emph{2 clicks}$. For a given quantum efficiency $\eta$ of the APD used for the detection of the two bins, the POVM elements of the TMD take the general form:
\begin{eqnarray}
\left\{
    \begin{array}{l}
       \hat{\Pi}_{0}(\eta)=\sum_{k=0}^{\infty}r_{k,0}(\eta) |k\rangle\langle k|  \\
       \\
       \hat{\Pi}_{1}(\eta)=\sum_{k=0}^{\infty} r_{k,1}(\eta) |k\rangle\langle k|\\
       \\
       \hat{\Pi}_{2}(\eta)=1-\hat{\Pi}_{0}(\eta)-\hat{\Pi}_{1}(\eta).    
    \end{array}\right.
    \label{POVMTMD}
    \end{eqnarray}
The POVM coefficients are calculated following the definition given in (\ref{condprob}):
\begin{eqnarray}
r_{k,0}(\eta)=P(0|k)&=&\sum_{k=0}^{p}R^p(1-R)^{k-p} {k \choose p}  r_{k-p,\textrm{\emph{off}}}\,r_{p,\textrm{\emph{off}}}\nonumber\\
&=&(1-\eta)^k
\end{eqnarray}
and
\begin{eqnarray}
r_{k,1}(\eta)&=&P(1|k)=\sum_{k=0}^{p}R^p(1-R)^{k-p}{k \choose p}\\
&&\qquad\qquad\qquad\times (r_{k-p,\textrm{\emph{on}}}\,r_{p,\textrm{\emph{off}}}+r_{p,\textrm{\emph{on}}}\,r_{k-p,\textrm{\emph{off}}})\nonumber\\
&=&(1-\eta)^k\sum_{k=0}^{p}R^p(1-R)^{k-p}{k \choose p}\nonumber\\
&&\qquad\qquad\qquad\times(-2+(1-\eta)^{p-k}+(1-\eta)^{-p})\nonumber\\
&=&(1-\eta)^{k}\Bigl[-2+\bigl(1+\frac{R\eta}{1-\eta}\bigr)^k+\bigl(1+\frac{T\eta}{1-\eta}\bigr)^k\Bigr].  \nonumber
\end{eqnarray}

\subsubsection{Including dark noise}

We now include the effect of Poissonian dark counts in the description. We note $\nu$ the mean number of dark counts in the detection window, which is defined before the `black-box' detector: $e^{-\nu}$ is thus the probability of no dark counts whatever the specific detector implementation. For instance for the TMD, where the mode is split into two temporal bins, $e^{-\nu/2}$ is the probability of no dark counts into each bin and $(e^{-\nu/2})^2=e^{-\nu}$ for the two bins. 

The general expression of the POVM in the presence of noise can be written as a function of the POVM without noise. Using (\ref{condprob}), one obtain for the APD
\begin{eqnarray}
\left\{
    \begin{array}{l}
r_{k,\textrm{\emph{off}}}(\eta,\nu)=e^{-\nu}r_{k,\textrm{\emph{off}}}(\eta) \\
r_{k,\textrm{\emph{on}}}(\eta,\nu)=1-e^{-\nu}r_{k,\textrm{\emph{off}}}(\eta)
    \end{array}\right.
   \label{POVMAPDNOISE}
    \end{eqnarray}
while for the TMD it can be read as:
\begin{eqnarray}
\left\{
    \begin{array}{l}
r_{k,0}(\eta,\nu)=e^{-\nu}r_{k,0}(\eta) \\
r_{k,1}(\eta,\nu)=e^{-\nu/2}r_{k,1}(\eta)+2(e^{-\nu/2}-e^{-\nu})r_{k,0}(\eta).
    \end{array}\right.
\label{POVMTMDNOISE}
    \end{eqnarray}

\subsection{Conditional state preparation with twin beams}

We now detail the conditional preparation. As explained before, a general strategy for the conditional preparation of quantum states relies on the use of entanglement: due to quantum correlations, measuring one mode of an entangled state results in projecting the other mode according to the result of this measurement. By denoting $\hat{\rho}_{AB}$ the two-mode state, a measurement performed on mode B characterized by the POVM $\hat{\Pi}_n$ leads to the preparation on mode A of the conditional state $\hat{\rho}_{c}$, which can be written as:
\begin{equation}
\hat{\rho}_{c}=\frac{1}{Pr(n)}\textrm{Tr}_{B}\left[\hat{\rho}_{AB}\,\hat{1}_A\otimes\hat{\Pi}_n\right]
\label{general}
\end{equation} 
where $\textrm{Tr}_{B}$ stands for the partial trace over mode B and $\hat{1}_A$ is the identity operator acting on mode A. $Pr(n)$ is the probability to obtain the outcome ``\emph{n}'' on mode B and is thus proportional to the preparation rate. It can be read as:
\begin{equation}
Pr(n)=\textrm{Tr}_{AB}\left[ \hat{\rho}_{AB}\,\hat{1}_A\otimes\hat{\Pi}_n\right].
\label{rate}
\end{equation} 

In this paper, we consider as an entangled state resource the twin beams produced by perfect two-mode spontaneous parametric down-conversion (SPDC):
 \begin{equation}
|\Psi_{AB}\rangle=\sqrt{1-|\lambda|^2}\,\sum_{k=0}^{\infty}\lambda^k|k\rangle_A|k\rangle_B
\label{psi}
\end{equation}
where the squeezing parameter $\lambda$, which varies between 0 and 1, is related to the intensity gain in the down-conversion process. This parameter determines the average photon number per mode as $\bar{N}={|\lambda|^2}/{(1-|\lambda|^2)}$. From the previous general expressions (\ref{general}) and (\ref{rate}), it follows that the conditional state $\hat{\rho}_c$ is given by
\begin{equation}
\hat{\rho}_{c}=\frac{\sum_{k=0}^{\infty}{|\lambda|^{2k}\,r_{k,n}\,|k\rangle\langle k |}}{\sum_{k=0}^{\infty}{|\lambda|^{2k}\,r_{k,n}}}
\label{phi}
\end{equation} 
where $r_{k,n}$ are the coefficients of the POVM element $\hat{\Pi}_{n}$. The probability $Pr(n)$ can also be written as a function of the squeezing parameter and POVM coefficients as
\begin{equation}
Pr(n)=(1-|\lambda|^2)\sum_{k=0}^{\infty}|\lambda|^{2k}\,r_{k,n}.
\end{equation}

For an ideal photon-number resolving detector, $r_{k,n}(\eta,\nu)=\delta_{k,n}$, the conditional state resulting from a heralding measurement $\hat{\Pi}_n=|n\rangle\langle n|$ reduces thus to a Fock state $|n\rangle$ for any values of $\lambda$ as expected. Only the preparation rate varies with this parameter. In the more general case, $\hat{\rho}_{c}$ tends to the normalized POVM element in the limit of high gain:
\begin{equation}
\hat{\rho}_{c} \overset{\lambda\rightarrow 1}{\longrightarrow} \frac{\hat{\Pi}_n}{\textrm{Tr}\left[\hat{\Pi}_n\right]}.
\label{state}
\end{equation}

Note that defining the non-classical character of a quantum measurement is not straightforward as the POVM elements are not normalized and thus do not correspond to a quantum state. However, conditional preparation provides a good insight into this subtle question. Given equation (\ref{state}), the non-classical properties of a detector can be seen as the non-classical properties of the state generated by a conditional measurement in the limit of high gain. Significantly, this expression also corresponds to the expression of the state retrodicted from the measurement outcome \cite{Barnett,Barnett2,Amri}. This state enables to make predictions (called retrodictions) about state preparation leading to this result. The non-classical properties of the detector can then be associated in this approach with the properties of the retrodicted state \cite{Amri}. 

\section{Quantum detector tomography and experimental determination of the POVM}
\label{section3}

We now turn to the experimental characterization of detectors. The complete characterization of a detector requires, for each of its POVM elements $\hat{\Pi}_{n}$, the knowledge of all the coefficients $\{r_{k,n}(\eta,\nu)\}$. This knowledge can be obtained experimentally by performing the so-called quantum detector tomography (QDT) \cite{Dauria,Luis,Fiurasek,Lundeen,Ronge09,Paris12,Brida12,Oxford12}. We present here the experimental setup for QDT and the results for the two detectors we are focusing on in the present study.

\subsection{Procedure and setup}
The procedure for QDT consists in probing the detector by sending a tomographically complete set of known states \{$\hat{\rho}_{j}$\} and in recording the number of times ($f_{n,j}$) each outcome ``\emph{n}'' occurred in the measurement for each probe state. By using equation ($\ref{Def}$), a maximum likelihood algorithm (ML) allows for reconstructing the POVM elements $\{\hat{\Pi}_{n}\}$ by searching for the set of the $r_{k,n}(\eta,\nu)$ coefficients that provide the optimal matching between the measured $f_{n,j}$ and the theoretical $p_{n,j}$. Significantly, the ML does not require any $\emph{a priori}$ assumptions on the detector under investigation. The details of the procedure can be found in \cite{Fiurasek}.

In our experiment, we used as probe states a set of coherent states $|\alpha_{j}\rangle\langle\alpha_{j}|$ of different amplitudes, generated by strongly attenuating the light coming from a pulsed laser source. The experimental setup is shown in Figure 1. Light pulses at 795 nm are provided by a femtosecond laser source (MIRA900) with a repetition rate set to 1.187 MHz thanks to a pulse picker device (9200 Coherent). In order to tune the average photon number per pulse, $|\alpha_{j}|^{2}$, the laser beam is sent through a variable attenuation device consisting of a half wave plate ($\lambda/2$) and a polarizing beam splitter ($PBS$). The power of the output beam is controlled by tapping a known fraction toward a power meter. A set of well calibrated optical densities (total \emph{O.D.$\approx$7}) supplies the attenuation necessary to reach the regime of few photons per pulse: $|\alpha_{j}|^{2}$ can be indeed adjusted from 0 to 100 by steps of 0.1. The probe beam is sent to the "black-box" containing the unknown detector through an optical fiber. The losses due to the coupling are taken into account when estimating the average photon number. The error on the average photon number is of the order of 5\%, mostly due to the residual laser intensity fluctuations between subsequent pulses.

\begin{figure*}
\begin{center}
\includegraphics[width=1.6\columnwidth]{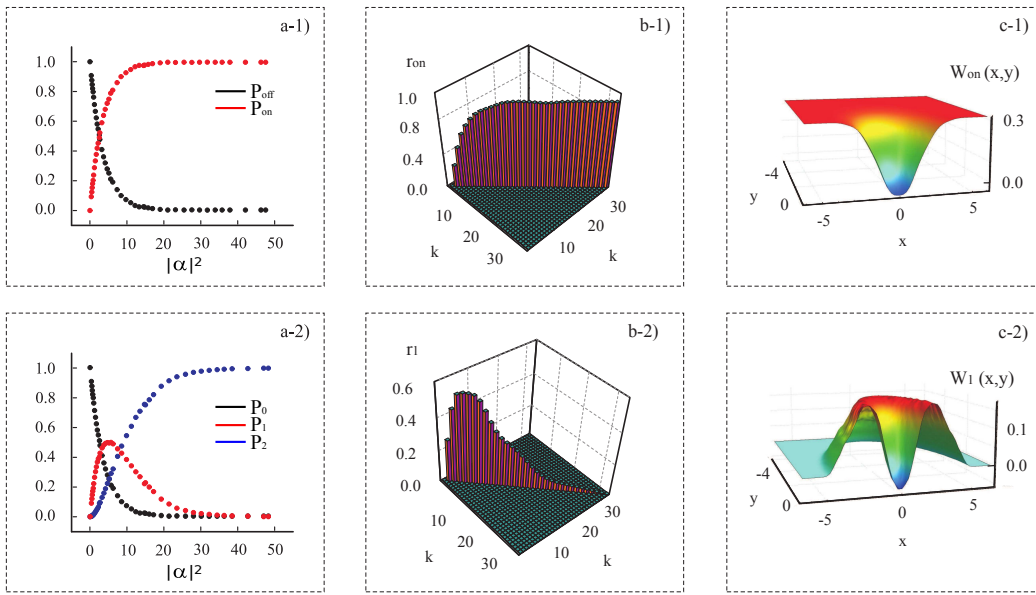}
\caption{Experimental data and reconstruction of the POVM elements. The first line corresponds to the APD while the second line gives the results for the TMD. Column (a) gives the bare tomographic data, i.e. the probability of the different outcomes as a function of the average photon number of the coherent probe pulses. Column (b) shows the reconstructed POVM $\hat{\Pi}_{\textrm{\emph{on}}}$ for the APD and $\hat{\Pi}_{1}$ for the TMD. Finally, Column (c) provides the Wigner representation of these operators.}
\label{figure2}
\end{center}
\end{figure*}

\subsection{Experimental reconstruction in the absence of noise}
We performed the detector tomography procedure on the two different devices presented in section \ref{section2}: a standard $\emph{on/off}$ avalanche photodiode (APD) and a single-loop time-multiplexed detector  (TMD). The APD (Perkin Elmer SPCM-AQR-13) is sketched on figure 1b  while the  multiplexed detector is shown in figure 1c. For the TMD, the light is split toward two channels: pulses going to the upper channel undergo a delay of $\Delta T=150$ ns with respect to those in the other channel. The two paths are eventually spatially recombined using orthogonal polarizations (this recombination avoids extra 50\% losses) and sent to an APD. As $\Delta T$ is larger than the detector dead time, the arrival of two photons will correspond to two subsequent events.  This setup leads to three possible responses: $\emph{off}$, $\emph{1 click}$ or $\emph{2 clicks}$. For both detectors, the overall quantum efficiency has been set to $0.28\pm0.02$. 

Figure 2 gives the experimental results of the QDT performed on both detectors (first line for the APD, second line for the TMD). Column (a) provides the frequency of the different outcomes as a function of the average photon number $|\alpha|^2$. Column (b) then gives the reconstructed operator, $\hat{\Pi}_{\textrm{\emph{on}}}$ for the APD and $\hat{\Pi}_{1}$ for the TMD, and column (c) finally shows the corresponding Wigner functions. As it can be seen, both are negative at the origin, which is a strong signature of the non-classicality of the detectors. Let us underline that this detector negativity is a critical property for quantum state engineering. Indeed, the preparation of a quantum state with negative Wigner function requires an heralding detector with negative Wigner function if the entangled resource has a positive one.

\begin{figure}[b!]
\begin{center}
\includegraphics[width=0.92\columnwidth]{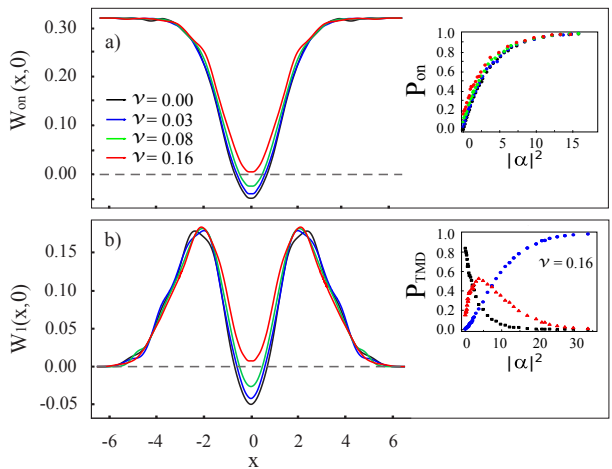}
\caption{Effect of dark counts. The cross-sections of the Wigner functions for the APD (a) and TMD (b) are given for different levels of dark counts. The inset of figure (a) gives the probability of the outcome 'on' for the different levels of noise while the inset of figure (b) provides the probability of the three possible outcomes (black=`no click', red=`1 click', blue=2 click') for a given value of the noise background.}
\label{figure3}
\end{center}
\end{figure}

\subsection{Experimental reconstruction in the presence of noise}
We performed similar QDT by adding noise to the detection process. The background noise is controlled by injecting on the detector, together with the probe states required for the QDT, a continuous-wave coherent beam at 1064 nm (Diabolo, Innolight). By adjusting the laser power, we are able to mimic different dark count levels with Poissonian statistics. As in section \ref{section2}, we note $\nu$ the mean number of dark counts in the detection window. Figure 3 gives the Wigner functions for the two devices and for different noise levels. As it can be seen, the negativity decreases with the noise, and eventually disappears. This gradual disappearance of the negativity witnesses in a quantitative way the degradation of the detector performances.

\section{Single-photon generation with APD and TMD}
\label{section4}
We now turn to the conditional generation of single-photon states using these two specific detectors. By simulating the entanglement resource, we first examine the state one would obtain using the characterized detectors. We then consider the more general case given by the theoretical POVM coefficients derived in section \ref{section2}, in particular for high quantum efficiency. This section illustrates the difference between the two detectors and the conditions in which the photon-number resolution ability of the two-channel TMD is useful. 

\subsection{Fidelity of the state generated with the characterized APD and TMD}

\begin{figure}
\begin{center}
\includegraphics[width=0.96\columnwidth]{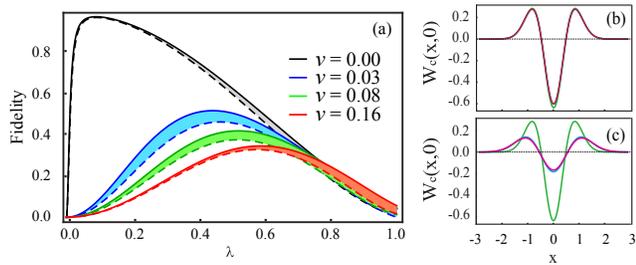}
\caption{Effect of the dark counts on the conditional state generation. The state is calculated from Eq.\ref{phi} by using the experimentally reconstructed POVM coefficients for the APD and for the TMD. Figure (a) gives the fidelity between a single-photon state and the state generated using the APD (dashed) or the TMD (straight), as a function of $\lambda$ and for different dark count levels. Figures (b) and (c) represent the cross-section of the Wigner functions of a single-photon state (green), of the states generated using an APD (red) and using the TMD (blue), in the absence of noise ($\nu=0$) and respectively for $\lambda=0.1$ and $\lambda=0.7$}
\label{figure4}
\end{center}
\end{figure}

Given the reconstructed POVM elements for the APD and the TMD, expression (\ref{phi}) enables us to determine the state one would conditionally generate by using these detectors and correlated photon pairs generated by SPDC. The fidelity of the generated state with the target one (pure state) can then be calculated as the overlap of the Wigner function, $W_c(x,y)$ associated to $\hat{\rho}_c$ and the Wigner function of the target state $W_t(x,y)$ \cite{Fiurasek2002,Jeong2007}:
\begin{equation}
F=\frac{\int_{-\infty}^{+\infty}W_c(x,y) W_t(x,y)dxdy}{\int_{-\infty}^{+\infty}|W_t|^2(x,y)dxdy}
\label{fidelity}
\end{equation}
In the following we calculate the fidelity with a single-photon state, whose Wigner function can be written as $W_t(x,y)=\frac{2}{\pi}e^{-2(x^2+y^2)}[-1+4(x^2+y^2)]$. 

Figure \ref{figure4}a gives the fidelity of the generated state for the APD and the TMD as a function of the squeezing parameter $\lambda$. We recall that increasing $\lambda$ leads to a higher number of photons in the twin modes. Therefore, in the limit of $\lambda \rightarrow 1$, the fidelity goes to 0 due to multi-photon pairs, which can not be resolved by the present detectors. Very low squeezing parameter is usually required to overcome their limited photon-number resolution. In this case, the two detectors behaves similarly.  Figures \ref{figure4}b and \ref{figure4}c provide the cross-sections of the Wigner functions in the absence of noise and for two different values, $\lambda=0.1$ and $\lambda=0.7$. In both cases, the generated states are very close. In the first case, however, the fidelity with a single-photon state is high while it strongly decreased for higher values.

In all the configurations, the TMD enables to reach a slightly larger fidelity due its photon-number resolution ability (yet limited). This gain appears for increasing value of $\lambda$, while it is expected to disappear for vanishing $\lambda$ due to the the low-photon number occupancy of the modes. As it can be seen, the difference between the results one would obtain using the APD and a single-loop TMD is however small. This is due to the low quantum efficiency of the avalanche photodiode used here. In the following, we present the general case, without being limited by the efficiency $\eta$ and noise  $\nu$ imposed by a specific experimental implementation. 

\begin{figure}
\begin{center}
\includegraphics[width=0.96\columnwidth]{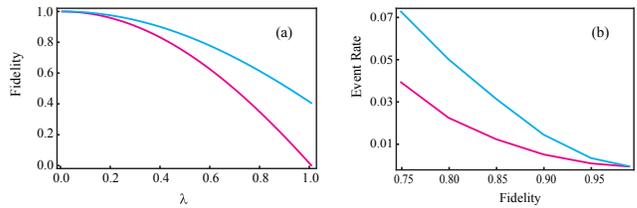}
\caption{Fidelity and event rate in the absence of noise ($\nu$=0) and for a quantum efficiency $\eta$=0.9. Plot (a) gives the fidelity between a single-photon state and the state generated using the APD (red) or the TMD (blue) as a function of the squeezing parameter $\lambda$. Plot (b) shows the event rate $R$ as a function of the targeted fidelity ($\lambda$ is chosen accordingly).}
\label{figure5}
\end{center}
\end{figure}

\subsection{General case}
By using the theoretical expressions for the POVM coefficients as given in section 2, we can predict the modification of the conditional states and of their generation probability (the rate of heralded events) as functions of the detector parameters.

In an ideal measurement, involving a projective detector the result 1-click of a measurement performed on mode B heralds the generation of a single-photon state in mode A  ($F=$1) , whatever the statistics of the initial twin beams, i.e. for all values of $\lambda$. In this case, the squeezing parameter $\lambda$ only affects the probability of generating the state, $R_{ideal}=(1-|\lambda|^2)|\lambda|^2$. 
For realistic detectors, the detector parameters affect both the generated state and the event rate. We compute here the fidelity of the generated state with a single-photon state, by making use of the expressions of $r_{k,\textrm{\emph{on}}}$ and $r_{k,1}$ of equations (\ref{POVMAPDNOISE}-\ref{POVMTMDNOISE}). For the APD, the fidelity $F_{\textrm{\emph{on}}}$ can be written as
\begin{equation}
F_{\textrm{\emph{on}}}=\frac{\lambda^2\cdot(1-e^{-\nu}(1-\eta))}{\frac{1}{1-\lambda^2}-\frac{e^{-\nu}}{1-(1-\eta)\lambda^2}}
\label{FidAPD}
\end{equation}
and the rate of events $R_{\textrm{\emph{on}}}$ is given by:
\begin{equation}
R_{\textrm{\emph{on}}}=1-\frac{e^{-\nu}(1-\lambda^2)}{1-(1-\eta)\lambda^2}.
\label{probAPD}
\end{equation}
Similarly, for the state heralded by the TMD, we obtain a fidelity
\begin{equation}
F_{1} =\frac{\lambda^2\cdot(1-\eta/2-e^{-\nu/2}(1-\eta))}{\frac{1}{1-(1-\eta/2)\lambda^2}-\frac{e^{-\nu/2}}{1-(1-\eta)\lambda^2}}
\label{FidTMD}
\end{equation}
with a corresponding event rate $R_{1}$:
\begin{equation}
R_{1}=2(1-\lambda^2)\left(\frac{-e^{-\nu}}{1-(1-\eta)\lambda^2}+\frac{-e^{-\nu/2}}{1-(1-\eta/2)\lambda^2}\right).
\label{proTMD}
\end{equation}
\begin{figure}[t!]
\begin{center}
\includegraphics[width=0.96\columnwidth]{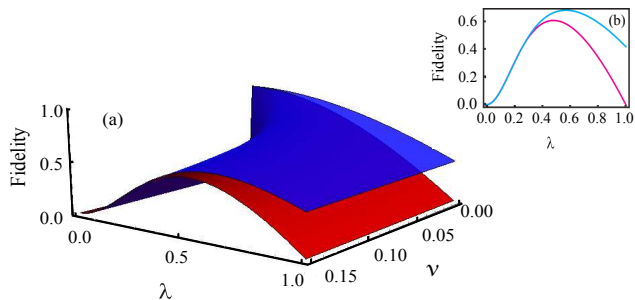}
\caption{Fidelity in the presence of noise and for a quantum efficiency $\eta$=0.9. Plot (a) gives the fidelity between a single-photon state and the state generated using the APD (red) or the TMD (blue) as a function of the squeezing parameter $\lambda$ and noise $\nu$. Plot (b) represents the fidelity as a function of $\lambda$ for a noise $\nu$=0.08.}
\label{figure6}
\end{center}
\end{figure}

As an example, we consider the case of a high quantum efficiency ($\eta=0.9$). In the absence of noise ($\nu=0$), figure \ref{figure5} gives the fidelity and preparation rate as a function of the squeezing parameter. As already seen in the previous section, the fidelity is decreasing with increasing $\lambda$ due to the multi-photon pairs emitted in the SPDC process. While the difference between the two detectors was small for a low quantum efficiency, it can be clearly seen now. The photon-number resolution ability of the TMD, yet limited, enables us to reach a larger fidelity for all values of $\lambda$. Similarly, for a targeted fidelity, the TMD also provides a higher count rate, as shown in figure \ref{figure5}b, as a higher $\lambda$ can be used. 

We finally compare the two detectors in the presence of noise. Figure \ref{figure6} shows the fidelity as a function of $\lambda$ and $\nu$, still for a quantum efficiency $\eta=0.9$. For a given value of noise, the TMD enables to reach a better fidelity. As the noise increases, the difference between the two detectors decreases. As it has been underlined previously, for a given noise level, there exists an optimal value of $\lambda$ for the fidelity. For the squeezing parameter going to zero, the fidelity drops as the signal to noise ratio decreases, i.e. the noise results in a large rate of false heralded events relative to the low signal. \\

\section{Conclusion}
Beyond its fundamental significance, the study of optical detectors has become in the last years more and more crucial for quantum information, in particular in the context of quantum state engineering and measurement-driven processes. We have presented here the general framework for conditional state preparation, which is an interplay between the resource and the heralding detector. In particular, we have taken the example of the conditional preparation of single-photon states and compared two detectors that can be used for such a preparation, an avalanche photodiode and a single-loop time-multiplexed detector. Four parameters are important in such a scheme: the entanglement resource defined here by the squeezing parameter $\lambda$, i.e. the photon-number occupancy of the correlated modes, the photon-number resolution ability of the detector, i.e. its capability to discriminate multi-photon events, the noise of the detector and its quantum efficiency. We investigated the heralded state one would obtain given different sets of these parameters. For low quantum efficiency, as we shown experimentally, the TMD only provides a small gain in the heralding process. However, the effect of the photon-number resolution ability of the TMD clearly appears for larger quantum efficiency, enabling in particular a higher preparation rate for a given fidelity of the generated state.\\

This work was supported by the BQR from Universit\'{e} Pierre et Marie Curie and the CHIST-ERA ERA-Net under the QScale project. V. D'Auria acknowledges the financial support from the European Commission under the Marie Curie Program. Claude Fabre and Julien Laurat are members of the Institut Universitaire de France.

\end{document}